\documentclass{emulateapj}
\usepackage{apjfonts}

\newcommand{\sfr}{M$_{\odot}$yr$^{-1}$}
\newcommand{\sfrd}{M$_{\odot}$yr$^{-1}$Mpc$^{-3}$}
\newcommand{\sfrv}{M$_{\odot}$yr$^{-1}$(L$_{\odot}{\times}10^{-10})^{-1}$}

\shorttitle{Spitzer observations of DRGs in the HDFS}
\shortauthors{Webb et al.}

\begin{document}


\title{Star Formation in  Distant Red Galaxies: Spitzer observations in the  Hubble Deep Field South}


\author{Tracy M.A. Webb\altaffilmark{1}}

\email{webb@strw.leidenuniv.nl}
\author{Pieter van Dokkum\altaffilmark{2,3}}
\author{Eiichi Egami \altaffilmark{4}}

\author{Giovanni Fazio\altaffilmark{5}}

\author{Marijn Franx\altaffilmark{1}}
\author{Eric Gawiser\altaffilmark{2,3,6}}
\author{David Herrera\altaffilmark{2,3}}

\author{Jiasheng Huang\altaffilmark{5}}

\author{Ivo Labbe\altaffilmark{8}}
\author{Paulina Lira\altaffilmark{6}}
\author{Danilo Marchesini\altaffilmark{2,3}}

\author{Jos\'{e} Maza\altaffilmark{6}}

\author{Ryan Quadri\altaffilmark{2}}

\author{Gregory Rudnick\altaffilmark{9}}
\author{Paul van der Werf\altaffilmark{1}}

\altaffiltext{1}{Leiden Observatory,  PO Box 9513,  2300 RA Leiden, The Netherlands}
\altaffiltext{2}{Department of Astronomy, Yale University, P.O. Box 208101, New Haven, CT 06520-8101}
\altaffiltext{3}{Yale Center for Astronomy \& Astrophysics, Yale University, P.O. Box 208121, New Haven, CT 06520}
\altaffiltext{4}{The University of Arizona, Steward Observatory, 933 N. Cherry Ave., Tuscon, AZ 85721 }
\altaffiltext{5}{CfA, 60 Garden Street, Cambridge, MA 02138}

\altaffiltext{6}{Departamento de Astronomia, Universidad de Chile, Casilla, 36-D, Santiago, Chile}


\altaffiltext{8}{OCIW, 813 Santa Barbara Street, Pasadena, CA, 91101}
\altaffiltext{9}{NOAO, 950 N. Cherry Ave., Tuscon, AZ, 85719}







\begin{abstract}
We present Spitzer 24{\micron} imaging of  1.5 $<z<$ 2.5 Distant Red Galaxies (DRGs) in the 10{\arcmin}$\times$10{\arcmin} Extended Hubble Deep Field South of the  Multiwavelength Survey by Yale-Chile. We detect 65\% of the DRGs with $K_\mathrm{AB} < $ 23.2 mag at S$_\mathrm{24{\mu}m}\gtrsim$ 40$\mu$Jy,  and conclude that the bulk of the DRG population are dusty active galaxies. A mid-infrared (MIR) color analysis with IRAC data suggests that the MIR fluxes are not dominated by buried AGN, and we interpret the high detection rate as evidence for a high average star formation rate of  $<\mathrm{SFR}>$ = 130$\pm$30 M$_{\odot}$yr$^{-1}$.  From this, we infer that DRGs are important contributors to the cosmic star formation rate density at $z\sim$ 2, at a level of $\sim$ 0.02 M$_{\odot}$yr$^{-1}$Mpc$^{-3}$ to our completeness limit of $K_\mathrm{AB}$ = 22.9 mag.
\end{abstract}





\section{Introduction}
Through a variety of observational methods we are building a census of the high-redshift universe, and can now directly study galaxies in the process of forming. Color selection criteria effectively select large samples of galaxies at $z>$ 1 such as  Lyman-break galaxies (LBGs) \citep{ste96}, and the massive Distant Red Galaxies (DRGs) \citep{fra03}. Selected by rest-frame optical colors to lie at $z\gtrsim$ 2, DRGs exhibit properties of both passively evolved stellar populations and dusty starburst galaxies \citep{for04,lab05,pap05}.   Characterizing  this complex population and placing it into the context of the  star formation history of the universe  will elucidate our  overall understanding of the assembly of massive galaxies.

Recent advances in the capabilities of  infrared facilities have opened a new window onto the high-redshift universe. Through the direct detection of  dust enshrouded activity, mid- and far-infrared (M/FIR) studies provide orthogonal information  to that gathered in the ultraviolet (UV), optical and near-infrared (NIR), which can be heavily biased by dust extinction. The work presented here addresses the nature of $z\sim$ 2 DRGs  through  {\it Spitzer} MIPS-24{\micron} and IRAC imaging.   We assume a ${\Omega}_\mathrm{M}$ = 0.3, ${\Omega}_{\Lambda}$ =  0.7 cosmology and H$_\circ$ = 70 km s$^{-1}$ Mpc$^{-1}$ throughout.

\section{The Sample, Spitzer Observations, and Photometry}
The DRG sample was drawn from the  10$\arcmin\times$10$\arcmin$  Extended Hubble Deep Field South (EHDF-S) of the Multiwavelength Survey by Yale-Chile (MUSYC) \citep{gaw05}.  DRGs are defined by  $(J-K)_{\mathrm{vega}}>$ 2.3 mag \citep{fra03,van03}, and, in this paper, we focus on DRGs with $K_{\mathrm{AB}}<$ 23.2 mag (total magnitudes are given throughout),  corresponding to the MUSYC 50\% completeness  limit (Quadri et al.,~in preparation).

Photometric redshifts  were derived  from $UBVRIz'JHK$ photometry using the code
presented in \citet{rud01,rud03}, through linear combinations of galaxy templates, with an  accuracy of  $\left\langle~|z_{spec} -
z_{phot}|~/~(1+z_{spec})~\right\rangle=0.05$ for $z>1.5$. Here, we  have restricted our analysis to redshifts 1.5 $<z<$ 2.5 where the 6.2{\micron}, 7.7{\micron}, and 8.6{\micron} Polycyclic Aromatic Hydrocarbon (PAH) features  of star-forming galaxies fall  into the Spitzer 24{\micron} filter.  This provides a sample of 79 DRGs  with a median redshift of $z$ = 2.0.


The 24$\mu$m observations were taken in the MIPS photometry mode and consist of 6 separate pointings.  The central deepest pointing represents $\sim$1 hour of frame time, and the 5 flanking pointings have 35 minutes of frame time each.   The raw data were reduced and combined with the Data Analysis Tool
developed by the MIPS instrument team \citep{gor05}. The final image covers $\sim$104 arcmin$^2$.

Imaging at 3.0$\mu$m, 4.5$\mu$m, 5.8$\mu$m and 8.0$\mu$m was taken in the IRAC mapping mode and mosaicked to produce a $\sim$140 arcmin$^2$ image.  Individual  frames of 200 seconds were combined for a final total frame time, per location on the map, of 20 minutes. The Spitzer Science Center provides a Basic Calibrated Data product with flat-field corrections, dark subtraction, and linearity and flux
calibrations. Additional steps conducted with the IRAC team customized software included pointing
refinement, distortion correction, and mosaicking \citep{hua04}.

Photometry on the 24{\micron} image was performed using the point spread function (PSF) fitting program DAOPHOT \citep{ste87}. The PSF was characterized using bright isolated point sources and then used to iteratively fit and subtract  each object in the image.  A small fraction (10/79) DRGs were highly confused with neighboring objects and these were removed  from further analysis.  We reached an average rms depth of $\sim$13$\mu$Jy, and found the source counts to be in good agreement with \citet{pap04}. Because not all the DRGs are detected at  24{\micron}, their flux densities and  upper limits  were determined in a consistent manner by  performing aperture photometry 
at each $K$-determined position after first subtracting all other objects in the catalog.  
We used 6{\arcsec} diameter apertures, and used the PSF stars to apply an aperture correction factor to total flux densities at 40{\arcsec}.

  Due to the smaller beam size at the shorter IRAC wavelengths ($<$ 2{\arcsec}), and because highly confused DRGs have been removed from the sample,  PSF fitting was not employed for the IRAC measurements.  Aperture photometry (3{\arcsec} diameter) was performed at the location of each DRG for each IRAC channel, again using isolated point sources to determine aperture corrections to total fluxes at 12.2$\arcsec$. The average rms depths are 2.6$\mu$Jy, 2.7$\mu$Jy, 4.3$\mu$Jy and 4.0$\mu$Jy at 3.6{\micron}, 4.5{\micron}, 5.8{\micron}, and 8.0{\micron}, respectively.

\section{MIR constraints on the nature of DRGs}
\begin{figure}
\epsscale{.95}
\hspace{0.75cm}\plotone{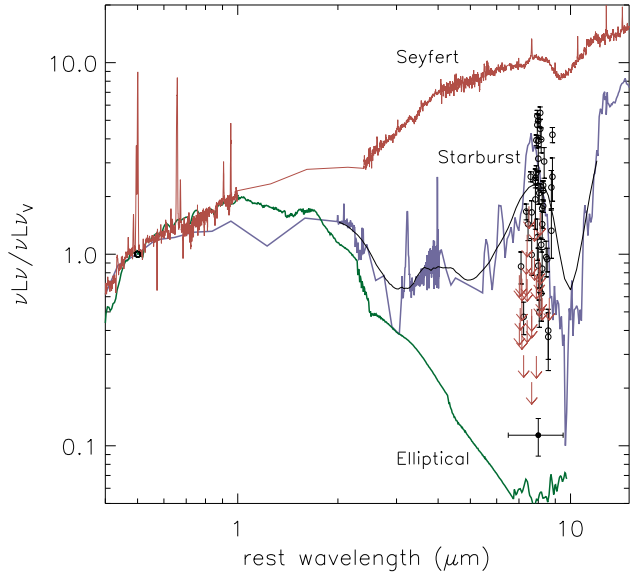}
\vspace*{1cm}
\caption{The 24{\micron} fluxes or limits of the DRGs, normalized to their rest-frame L$_\mathrm{v}$.  Points are DRGs with 24{\micron} detections, and  arrows show  3$\sigma$ upper-limits for the remainder. The single solid point below the arrows denotes the stacked measurement of DRGs which are not individually detected.  Three SEDs are shown: the green line corresponds to an elliptical galaxy model  \citep{cww} extrapolated to longer wavelengths using ISO observations of local ellipticals;  the blue line to the observed SED of Arp220; and the red line to the observed SED of NGC1068 (Jiasheng Huang, personal communication). Because of the rapid variation in flux with wavelength for the Arp 220 SED in the MIR we also show this SED smoothed by the 24{\micron} filter transmission curve (smooth black line).  \label{sed}}
\end{figure}

At 1.5 $<z<$ 2.5, the 24{\micron} filter samples $\sim$ 6-10{\micron} in the rest-frame and the 24{\micron} flux offers a powerful method of differentiating between two basic SED types: active dusty galaxies, i.e., those that are powered by either star formation or active galactic nuclei (AGN),  both of which  produce substantial mid-infrared emission, and passively evolved systems whose stellar flux drops drastically longward of $\sim$2{\micron} (see Fig.~\ref{sed}).   In  pure starburst galaxies, the MIR emission is dominated by PAH features which are strong relative to the underlying dust continuum, which only begins to rise above $\sim$ 10{\micron}. The hard radiation field  of an AGN destroys  PAH carriers  and the   continuum emission from hot small dust grains is strong throughout the MIR \citep{gen00}.

In Fig.~\ref{sed} we show the 24{\micron} flux densities for the DRGs, normalized to their rest-frame L$_{\mathrm{V}}$, and overlaid with three template SEDs; 65\% are detected  at 24{\micron} and the remainder are shown as 3$\sigma$ upper limits. This detection rate is in line with that found by \citep{pap05} for DRGs the GOODS Chandra Deep Field South, to a similar depth.    All of the 24{\micron}-detected DRGs are incompatible with old stellar populations and dusty starburst or AGN galaxies are required to produce such strong mid-IR emission.  The sample shows a wide range in 24{\micron} flux which we interpret primarily as a range in star formation rates. However,   a real difference in PAH strengths, relative to the continuum,  due to a range in metallicity or the hardness of the radiation field \citep[e.g.,][]{eng05,hog05} could also be important, as could a variation in the level of AGN contamination of the MIR flux, though as we discuss below we deem this latter possibility unlikely.  

The SEDs of  DRGs which are not detected at 24{\micron} (35\%)   are consistent with passively evolved systems,  but also lower luminosity or less dusty starbursts, or starbursts which have weak or absent PAH signatures as explained above.    To investigate the nature of the 24{\micron}-faint DRGs we used a  stacking analysis to measure their mean 24{\micron} emission and find  $S_{24{\mu}\mathrm{m}}$ = 10 $\pm$ 2 $\mu$Jy.  Thus, the remaining 35\% of the sample cannot be entirely composed of passively evolved systems  but  contains some fraction of dusty starburst or AGN galaxies.  A bootstrapping analysis indicates that this measurement is not dominated by a small number of systems, but beyond this we cannot constrain the fraction of starburst or AGN  systems within this group. 

\begin{figure}
\hspace{0.5cm}\plotone{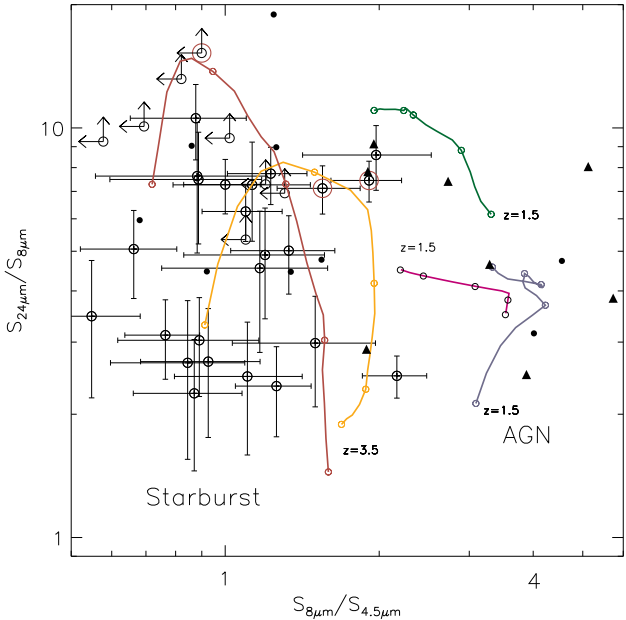}
\vspace*{1cm}
\caption{A MIR color-color plot.  The  tracks correspond to two starburst SEDs (left-most, orange-Arp220; red-M82) and three AGN dominated SEDs (right-most, magenta-MRK231; purple-Seyfert 1 NGC5506; green-Seyfert 2 NGC1068)\label{ratio}. The SED tracks begin at $z=$ 1.5 and are marked by circles in $z=$ 0.5 steps, to $z=$ 3.5.   The open points show DRGs from this work which are detected at 24{\micron}, with the radio detected objects indicated by the large red circles.    Also shown are  Spitzer observed SMGs \citep[solid circles;][]{ega04} and  optically-faint AGN \citep[solid triangles;][]{rig05}. \label{col}   }
\end{figure}

Combining the 24{\micron} measurements with IRAC data provides  a  color diagnostic of AGN activity (Fig.~\ref{col}) \citep{ivi04,ega04,hua05}.  This technique  relies on the different rest-frame NIR SED properties of AGN and starburst galaxies:  stellar emission in starburst galaxies leads to a relatively flat SED in the rest-frame NIR,  while hot dust around an AGN produces a power-law that increses toward longer wavelengths.  We have detected 50\% of the 24{\micron}-bright DRGs at 4.5{\micron} and 8{\micron}, another 20\% at 4.5{\micron}, but not 8.0{\micron}, 5\% in neither, and 25\% fall off the IRAC channel 2/4 field-of-view.  The colors (Fig.~\ref{col})   indicate that  star formation dominates the MIR emission of the DRGs,  but  as a  conservative cut we consider the four objets with $S_\mathrm{8{\mu}m}/S_\mathrm{4.5{\mu}m}>$ 1.5  to have {\it possible }   AGN contamination.   This is in agreement with results from the X-ray \citep{rub04,red05} and optical \citep{van03} which  also indicate that AGN do not significantly contribute to the observed luminosity of DRGs.  It should be remembered, however, that very heavily obscured or weak AGN, though energetically unimportant in terms of observed MIR properties, may still be present.

\section{ Star Formation Rates of DRGs}

PAH and MIR emission can be used to estimate the  current star formation rate (SFR) of a galaxy \citep{wu05}.  Using an Arp220  SED template  we extrapolated from 24{\micron}/(1+$z$) to rest-frame 6.75{\micron}. We then estimated L$_\mathrm{IR}$ through the  6.75{\micron}-F$_\mathrm{FIR}$ relation calibrated  locally with the {\it Infrared Space Observatory}, which shows a scatter of  $\sim$50\% \citep{elb02}.   This value can  then be converted to a SFR following the  L$_{\mathrm{IR}}$-SFR relationship of \citet{ken98} with an uncertainty of a factor $\sim$ 2-3.  We note that adopting an M82-like SED instead of Arp220 in the first step results in SFR estimates which are systematically 1.4$\times$ lower.

\begin{figure}
\epsscale{0.95}
\hspace{0.75cm}\plotone{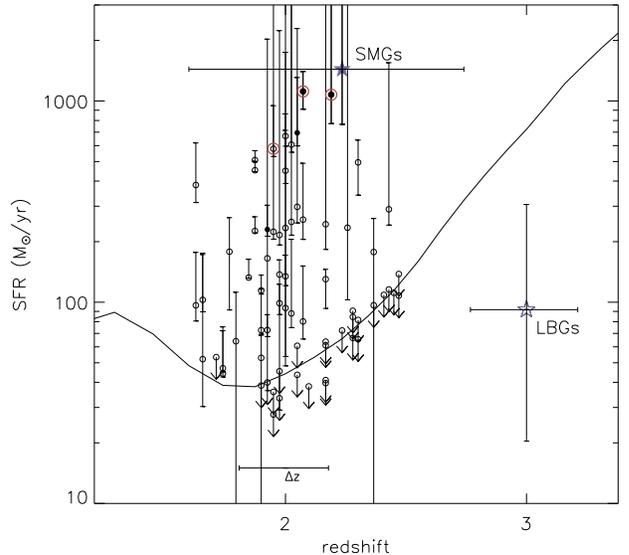}
\vspace*{1cm}
\caption{The SFRs for  DRGs as a function of redshift (circles). Filled circles mark DRGs which may contain AGN, and red circles mark radio detected galaxies. The solid line corresponds to  the average effective SFR depth as a function of redshift, for these data. Also shown for reference is the median SFR for radio-detected submm galaxies from \citet{cha05} (solid star) and that of Lyman-break galaxies from \citet{sha01} (open star). \label{sfrz}}
\end{figure}

In Fig.~\ref{sfrz} we plot the 24{\micron} star formation rates or upper limits of the individual DRGs with  redshift.   The population is not homogeneous and individual galaxies range in SFR from $<$ 30 {\sfr} to $\sim$1000 {\sfr}.    A strong AGN contribution to the MIR flux will contaminate this estimate, by boosting the inferred SFR, and sources which may contain AGN are marked.  Also highlighted are three DRGs which are detected in the  deep radio map of \citet{huy05}, two of which are possible AGN based on their MIR colors.   Studies indicate that the fraction of AGN contamination in the $\mu$Jy radio population increases at higher redshift  \citep{ric00,gar01}, and thus, the radio detected DRGs may indeed contain weak AGN.   However, we note that these systems fall within 1$\sigma$ of the 
FIR/radio flux correlation of \citet{con92} (with no systematic offset), which only holds for starburst galaxies, and thus significant AGN contamination is unlikely.  

Using a  stacking analysis, we calculated an average SFR for the entire DRG sample of $<$SFR$>$ =  130$\pm$30 M$_{\odot}$yr$^{-1}$. This is the uncertainty-weighted mean  of {\it all} 69 DRGs, assuming no AGN contamination and a mean redshift of $z$ = 2, with the uncertainty in the mean estimated from a bootstrapping analysis. This result is in good agreement with the optical SED modeling results  of \citet{for04} of 120 {\sfr},  and X-ray and submillimeter stacking analyses which estimate  $\sim$100-200 {\sfr} and $\sim$100 {\sfr}, respectively \citep{rub04,red05,knu05}.

 A significant source of error in this calculation  is the photometric redshift uncertainty,  since the rest-frame flux  of starburst galaxies changes rapidly with wavelength in the region surrounding the PAH features.  This effect was quantified by recalculating SFRs for simulated redshifts within Gaussian confidence intervals, and the resulting uncertainties are shown in Fig.~\ref{sfrz}.  The uncertainties are substantial, ranging from a factor of two to almost an order of magnitude.

We  calculated the optically normalized star formation rates (SFR/L$_\mathrm{V}$) for the DRGs, which provide an indication of  their age and/or dust content.  We find $<$ SFR/L$_V>\sim$ 19\footnote{{\sfrv}}, which is in excellent agreement with the submillimeter study of DRGs by \citet{knu05}, who measure  $<$ SFR/L$_{\mathrm{V}}>\sim$  20. However, the values for the DRGs show  large scatter, ranging from   SFR/L$_{\mathrm{V}}<$ 5 to 200.  Monte Carlo simulations indicate that the redshift uncertainties cannot account for the full range of this scatter and, moreover, the three radio detected galaxies have the highest SFR/L$_{\mathrm{V}}$, for a given L$_{\mathrm{V}}$; as their radio flux confirms that they are extreme systems, their high SFR/L$_{\mathrm{V}}$ values are unlikely to stem solely from redshift uncertainties.
  
 \section{Contribution to the Cosmic Star Formation Density}

Using the 24{\micron} SFR  estimates presented here, and the number density of DRGs in the MUSYC fields (known to a factor of two; Quadri et al.,~in preparation),  we estimate a star formation rate density (SFRD) for the DRGs of $\sim$0.02 M$_{\odot}$yr$^{-1}$Mpc$^{-1}$ over the redshift range 1.5 $<z<$ 2.5, and to a depth of $K_\mathrm{AB}$ =  22.9 mag.    At this depth our sample is 100\% complete, and therefore requires no completeness correction.  A number of  uncertainties affect  this estimate, including those described in the previous sections:  errors in the  photometric redshifts, the uncertain conversion of an observed 24{\micron} flux  to a SFR, and the potential contamination from AGN.   Also important  are possible differences in dust properties and the Initial Mass Function from low to high redshift, which are difficult to quantify. 
Nevertheless,  though the uncertainties in the individual star formation rates are high, the average star formation rate of the population is known to better accuracy since the redshift uncertainties, at least, are largely statistical. 

The total SFRD, integrated over all galaxies,  provides a census of the star formation history of the universe, and the relative contributions of different galaxy populations is indicative of their importance in assembling stellar mass.  In a general sense, there is now reasonable agreement between IR and optical SFRD estimates at $z\sim$ 2. \citet{red05} estimate a SFRD (from X-ray measurements) of 0.10 {\sfrd} over 1.4 $< z<$ 2.6, for optical and NIR selected galaxies, with DRGs producing 20\% of this total. Likewise, \citet{cha05}  estimate $\sim$ 0.1 {\sfrd} for  submm-selected systems, after extending the submm luminosity function down to $S_{850{\mu}{\mathrm{m}}}$ = 1 mJy, and applying a correction for AGN contamination. Thus, the two techniques converge on a similar value, and assuming the $S_\mathrm{850{\mu}m}\sim$ 1 mJy population encompasses the bulk of the optical and NIR selected galaxies, and vice versa, this can be taken as a total SFRD at $z\sim$ 2  (but see both papers for a discussion of uncertainties and assumptions). 

 Our completeness limit of $K_\mathrm{AB}$ = 22.9 mag is roughly equal to the \citet{red05} DRG depth over a similar redshift range, and we find the same SFRD for DRGs of $\sim$ 0.02 {\sfrd};  thus, the X-ray and MIR determined SFRD measurements agree that the DRGs produce roughly 20\% of the total SFRD of optical and NIR selected star forming galaxies.  A consistent picture is seen at longer wavelengths.   If we use the 24{\micron} flux density of each DRG to predict its approximate 850{\micron} flux density, we find that our SFRD estimate is dominated by objects with $S_\mathrm{850{\mu}m}>$ 1mJy, and thus again  DRGs  produce $\sim$ 20\% of the total SFRD as measured at submm wavelengths. In conclusion, DRGs, to a depth of $K_\mathrm{AB}$ = 22.9 mag are important contributors to the total SFRD at $z\sim$ 2, and by extension, to the assembly of massive galaxies.

\section{Summary}

The analysis presented here indicate that DRGs constitute a heterogeneous population, with the majority consisting of luminous and dusty starburst galaxies. We find that $>$65\% of the E-HDFS sample are such systems and this is in good agreement with similar Spitzer studies \citep{pap05}, and  rest-frame optical SED modeling  \citep{for04,lab05}.  Using the 24{\micron} flux we estimate an average SFR for the population (to $K_\mathrm{AB}$ = 23.2) of 130$\pm$30 {\sfr},  in line with earlier estimates from  optical, submm and X-ray studies. Normalizing the SFR to the rest-frame L$_{\mathrm{V}}$ yields  SFR/L$_{\mathrm{V}}\sim$ 19 {\sfrv}, but with substantial scatter around this value.  The scatter cannot be explained solely by redshift uncertainties, and indicates  a real difference in the individual  properties  DRGs.   Overall, DRGs are important contributors to the SFRD at $z\sim$ 2 at a level of  0.02 {\sfrd} to our 100\% completeness limit of $K_\mathrm{AB}$ = 22.9.

 \acknowledgments
Research by TW is supported by a VENI Research Fellowship, through the Nederlanse Organisatie voor Wetenschappelijk Onderzoek. DM is supported by NASA LTSA NNG04GE12G.  We are grateful to N.M.~F\"orster Schreiber for a careful reading and insightful comments on an early version of this work.


\begin{thebibliography}{}
\bibitem[Chapman et al.(2005)]{cha05} Chapman, S. C, Blain, A.W., Smail, I., \& Ivison, R. J. 2005, \apj, 622, 772
\bibitem[Coleman, Wu, \& Weedman(1980)]{cww} Coleman, G. D., Wu, C. -C., \& Weedman, D. W. 1980, ApJS, 43, 393
\bibitem[Condon(1992)]{con92} Condon, J.J., 1992, ARA\&A, 30, 575 
\bibitem[Egami et al.(2004)]{ega04} Egami, E., et al. 2004, ApJS, 154, 130
\bibitem[Elbaz et al.(2002)]{elb02} Elbaz, D., Cesarsky, C. J., Chanial, P., Aussel, H., Franceschini, A., Fadda, D., \& Chary, R. R. 2002, A\&A, 384, 848
\bibitem[Engelbracht et al.(2005)]{eng05} Engelbracht, C. W., Gordon, K. D., Rieke, G. H., Werner, M. W., Dale, D. A., \& Latter, W. B. 2005, \apj, 628, L29
\bibitem[F\"orster Schreiber et al.(2004)]{for04} F\"orster Schreiber, N. M. et al., 2004, \apj, 616, 40
\bibitem[Franx et al.(2003)]{fra03} Franx, M. et al., 2003, \apj, 587, 79L 
\bibitem[Garrett(2001)]{gar01} Garret, M. A. 2001, in ASP Conf Ser.  249, The Central kpc of Starbursts and AGN, ed.~J. H. Knapen, J. E. Beckman, I. Shlosman, \& T. J. Mahoney,  652
\bibitem[Gawiser et al.(2005)]{gaw05} Gawiser, E. et al. 2005, submitted to ApJ, astro-ph/0509202
\bibitem[Genzel \& Cersarsky(2000)]{gen00} Genzel, R. \& Cesarsky, C. J. 2000, ARA\&A, 38, 761
\bibitem[Gordon et al.(2004)]{gor05} Gordon, K. D. et al. 2005, PASP, 117, 503 
\bibitem[Hogg et al.(2005)]{hog05} Hogg, D. W., Tremonti, C. A., Blanton, M. R., Finkbeiner, D. P., Padmanabhan, N., Quintero, A. D., Schlegel, D. J., \& Wherry, N. 2005, \apj, 624, 162
\bibitem[Huang et al.(2004)]{hua04} Huang, J. -S. et al. 2004, ApJS, 154, 44
\bibitem[Huang et al.(2005)]{hua05} Huang, J. et al., 2005, ApJ, in press
\bibitem[Huynh et al.(2005)]{huy05} Huynh, M., Jackson, C. A., Norris, R., \& Prandoni, I. 2005, AJ, 130, 1373
\bibitem[Ivison et al.(2004)]{ivi04} Ivison, R. J. et al. 2004, ApJS, 154, 124
\bibitem[Kennicutt (1998)]{ken98} Kennicutt, R. C., Jr. 1998, ApJ, 498, 541
\bibitem[Knudsen et al.(2005)]{knu05}  Knudsen, K. K. et al. 2005, submitted to ApJ
\bibitem[Labb\'e et al.(2005)]{lab05} Labb\'e, I. et al. 2005, \apj, 624, 81L
\bibitem[Papovich et al.(2004)]{pap04} Papovich, C. et al. 2004, ApJS, 154, 70
\bibitem[Papovich et al.(2005)]{pap05} Papovich, C. et al. 2005, \apj, in press
\bibitem[Reddy et al.(2005)]{red05} Reddy, N. A., Erb, D. K., Steidel, C. C., Shapley, A. E., Adelberger, K. L., \& Pettini, M. 2005, \apj, 633, 748
\bibitem[Richards(2000)]{ric00} Richards, E. A. 2000, \apj, 533, 611
\bibitem[Rigby et al.(2005)]{rig05} Rigby, J. R. et al. 2005, \apj, 627, 134
\bibitem[Rudnick et al.(2001)]{rud01} Rudnick, G. et al. 2001, AJ, 122, 2205
\bibitem[Rudnick et al.(2003)]{rud03} Rudnick, G. et al. 2003, \apj, 599, 847
\bibitem[Rubin et al.(2004)]{rub04} Rubin, K. H.R., van Dokkum, P. G., Coppi, P., Johnson, O., F\"orster Schreiber, N. L., Franx, M., \& van der Werf, P. 2004, \apj, 613, 5 
\bibitem[Shapley et al.(2001)]{sha01} Shapley, A. E. et al. 2001, ApJ 562, 95
\bibitem[Steidel et al.(1996)]{ste96} Steidel, C. C., Adelberger, K. L., Giavalisco, M., Dickinson, M., \& Pettini, M. 1996, \apj, 519, 1
\bibitem[Stetson(1987)]{ste87} Stetson, Peter, B. 1987, PASP, 99, 191
\bibitem[van Dokkum et al.(2003)]{van03} van Dokkum, P. G. et al. 2003, \apj, 587, 83L
\bibitem[Wu et al.(2005)]{wu05} Wu H. et al. 2005, \apj, 632, 79
\end{thebibliography}
\end{document}